\documentstyle[nato,epsf]{crckapb}
\begin{opening}
\title{MAGNETIC MONOPOLES AND VORTICES IN THE \protect\\ STANDARD MODEL OF
ELECTROWEAK INTERACTIONS}

\author{A. ACHUCARRO}
\institute{Department of Theoretical Physics\\ The  University of
the Basque Country\\ Apt 644, 48080 Bilbao, Spain\\
and\\
Institute for Theoretical Physics\\ University of Groningen\\
Nijenborgh 4, 9747 AG Groningen, The Netherlands}

\end{opening}

\runningtitle{MAGNETIC MONOPOLES}
\begin{document}

\def\eqalign#1{\null\,\vcenter{\openup1\jot \mathsurround=0pt
     \ialign{\strut\hfil$\displaystyle{##}$&$\displaystyle{{}##}$\hfil
      \crcr#1\crcr}}\,}
\def\be{\begin{equation}}
\def\ee{\end{equation}}

{\small \narrower
\noindent
These lectures start with an elementary introduction to the subject of
magnetic monopoles which should be accesible from any physics
background. In the Weinberg-Salam model of electroweak interactions,
magnetic monopoles appear at the ends of a type of non-topological
vortices called {\it electroweak strings}. These will also be
discussed, as well as recent simulations of their formation during a
phase transition which indicate that, in the (unphysical) range of
parameters in which the strings are classically stable, they can form
with a density comparable to topological vortices.

}

\section{Introduction}
Last year marked the 50th anniversary of one of P.A.M. Dirac's most
profound and famous papers - on magnetic monopoles. While his 1931
paper is usually considered the official birthday of magnetic
monopoles, his 1948 paper is where he really showed that it was
possible to have a consistent quantum theory of magnetic poles in
conjunction with electric charges, and described the interaction
between them \cite{D31}.

A standard reference on magnetic monopoles is Preskill's lectures in
the 1985 Les Houches school \cite{P85}.  Here I have tried to present
magnetic monopoles and vortices in a way that makes them accessible to
physicists who are not so familiar with the language of high energy
physics, in particular with {\it e.g.} non-abelian gauge theories. As
a result, these lectures are much less technical.

The theme of this school is the use of topological defects
as a tool to understand the dynamics of phase transitions out of
equilibrium.  It turns out that, since magnetic monopoles and
electroweak vortices are non-topological in the Weinberg-Salam model,
their study can be particularly interesting in order to understand the
role of gauge fields in defect formation during phase transitions.

\section{The elusive monopole}

The problem with magnetic monopoles is well known in the context of
Maxwell's equations. If the electromagnetic field is described by the
vector potential
$\vec A$, then
\be {\vec \nabla} \cdot {\vec B} = 0 \ee so
there can be no sources or sinks for the magnetic field.

It would be fair to say that the experimental evidence for the
existence of magnetic monopoles is not good.  The 1998 Review of Particle
Properties by the Particle Data Group \cite{PDG98}
shows the result of monopole searches in particle accelerators: not a
monopole in sight. On the other hand, cosmic ray searches have
essentially only one event for which there seems to be no alternative
explanation, observed by Cabrera in Stanford in 1982 \cite{C82}.  The Cabrera
detector, like many others that failed to find anything before or after it,
consisted of a superconducting ring where a persistent
current was monitored for a long time; in \cite{C82} the loop had an
area of $20 \  \rm cm^2$ and was monitored for a total of 151 days. During
this time a single event was recorded which could be ascribed to a
magnetically charged particle with one Dirac unit of magnetic
charge $q_m = 2\pi \hbar c / e$. A magnetic monopole.

Under the circumstances, the experimental evidence is neatly
summarized by the sentence (whose author is unfortunately unknown to me)
{\it ``It is not clear that nobody has ever seen a magnetic monopole; what
is clear is that nobody has ever seen two''}.
And yet since Dirac's seminal work there have been over three thousand
papers in the literature about magnetic monopoles! \cite{M98}

The reasons behind this fascination with monopoles have evolved with
time, but they are basically three:

- the existence of monopoles would explain the quantisation of
electric charge (for which there is no alternative explanation to this day).
In his 1948 paper Dirac
says: {\it ``If one supposes that a particle with a single magnetic pole
can exist and that it interacts with charged particles, the laws of
quantum mechanics lead to the requirement that the electric charges
shall be quantized -- all charges must be integral multiples of a unit
charge e connected with the pole strength by the formula $eg =
{1\over 2} \hbar c$. Since electric charges are known to be quantized
and no reason for this has yet been proposed apart from the existence
of magnetic poles, we have here a reason for taking magnetic monopoles
seriously''.} He then goes on to say that {the fact that they have not
yet been observed may be ascribed to the large value of the quantum of
the pole}.

- a large class of theories that include electromagnetism as a subset
predict magnetic monopoles as solitons, as was shown by 't Hooft and
Polyakov in 1974 \cite{HP74}, and

- if magnetic monopoles exist, Maxwell's equations are symmetric under
the exchange of electric and magnetic fields. This {\it duality}
symmetry relates small electric charge to large magnetic charge and
viceversa. A generalization of this symmetry to non-abelian theories
would mean that the dual theory (of weakly coupled monopoles) could be
used to understand strongly coupled non-abelian gauge theories and, in
particular, confinement, for which there is no other analytic
approach.
While the idea is not
new, some of the most important developments in this area are
fairly recent; but they fall outside the scope of these
lectures, and I refer the reader to an excellent review by Harvey
\cite{H96}.

In what follows I will take $c=\hbar=1$. I will also depart from
Dirac's notation and use $q_m$ to refer to the magnetic charge; $g$
will be the $SU(2)$ coupling constant.

One final comment. I think everyone attending this school is aware of
the language problems between high energy and condensed matter
physicists. A relatively common source of confusion is the use of
the word {\it gauge symmetry}, which can mean different things to the
two communities. We all agree that electromagnetism has a gauge
symmetry, it is the symmetry that allows {\it local} (that is,
position--dependent) changes in the phase of the wave function and a
compensating change of gauge in the vector potential
\be \eqalign{
\psi (t, \vec {x}) &\to e^{i e \chi(t, \vec{x})} \psi (t, \vec {x})
\cr A_\mu (t, \vec {x}) &\to A_\mu (t, \vec {x}) + {\nabla}_\mu \chi
(t, \vec {x}) \qquad \mu = 0,1,2,3  \ .\cr} \ee
The
electromagnetic tensor $F_{\mu\nu} \equiv \nabla_\mu A_\nu -
\nabla_\nu A_\mu $, also known as the {\it field strength}, is
unchanged by this transformation, while
the covariant derivative
of $\psi$, $~\nabla_\mu \psi - ieA_\mu \psi$, ~  transforms in the same way
as $\psi$ itself (thus the name {\it covariant}).

High energy physicists use the term ``gauge symmetry'' to indicate any
symmetry which is local, whether or not it corresponds to a $U(1)$
transformation. Most condensed matter physicists, on the other hand,
will talk about gauge symmetries to indicate a change in the phase of
the wave function (a $U(1)$ transformation), whether or not there are
vector potentials around. Thus, the transformation \be \Psi \to e^{i
\alpha} \Psi \qquad {\alpha = {\rm const}} \ee is a gauge
transformation in the condensed matter literature, but not in the high
energy literature (where it would be called a {\it global} $U(1)$
transformation). On the other hand a high energy physicist will talk
about {\it e.g.}  a $SU(2)$ or $SO(3)$ gauge transformation, meaning
what is best described as a ``position-dependent rotation'' in
internal space. Since there are three degrees of freedom associated
with rotations, we need three vector potentials
\be
\underline{W}_\mu
= (W_\mu^1,W_\mu^2,W_\mu^3)
\qquad
{\bf W }_\mu \equiv
\underline{W}_\mu \cdot \underline{\tau} = {1 \over 2} \pmatrix{ W_\mu^3
\qquad
W_\mu^1 - i W_\mu^2 \cr W_\mu^1 + i W_\mu^2 \qquad - W_\mu^3\cr} ,
\ee
where $\underline{\tau} = (\tau^1, \tau^2, \tau^3)$ are the Pauli
spin matrices. The transformation law for the gauge potentials is
\be
{\bf W}_\mu (t, \vec {x}) \to M^{-1}(t, \vec {x}) {\bf W}_\mu (t, \vec {x}) M
(t, \vec {x}) + {1 \over g} M^{-1} (t, \vec {x}) \nabla_\mu M (t, \vec {x}) \ee
where $M(t,\vec{x})$ is a $SU(2)$ matrix at each point in spacetime;
note that
if $M$ were independent of position, as in the case of spin, there
would be no need for gauge potentials and the symmetry would be called
a {\it global} $SU(2)$ symmetry.

The transformation is non-abelian and moreover the field strength,
which has to be generalized from $F_{\mu\nu} = \partial_\mu A_\nu -
\partial_\nu A_\mu $ to
\be {\bf G}_{\mu\nu} = \partial_\mu {\bf
W}_\nu - \partial_\nu {\bf W}_\mu + g [{\bf W}_\mu {\bf W}_\nu - {\bf
W}_\nu {\bf W}_\mu ]
\ee
or
\be
{\underline G}_{\mu\nu} = \partial_\mu
{\underline W}_\nu - \partial_\nu {\underline W}_\mu + g {\underline
W}_\mu \times {\underline W}_\nu \ee
(with the cross product taken in internal space)
is no longer invariant but changes
like
\be \underline{G}_{\mu\nu} \cdot \underline{\tau} \to
M^{-1}(t, \vec {x})~\underline {G}_{\mu\nu}\cdot\underline{\tau}~ M (t,
\vec {x})
\ee

The transformation law of the scalars depends on the group
representation to which they belong and we list here two that will be
relevant later. The fundamental representation of $SU(2)$ is a doublet
of complex fields $\Phi = (\phi_1, \phi_2)$ whose transformation law
and covariant derivative are \be \Phi \to M^{-1}(t, \vec{x}) \Phi
\qquad {\rm and}\qquad D_\mu \Phi \equiv \nabla_\mu {\Phi} + g {\bf
W}_\mu {\Phi} \qquad {\rm respectively.}\ee The adjoint representation
is a triplet of real scalars $\underline{\phi} =(\phi^1,\phi^2,\phi^3
)$ which, like the gauge potentials, can be assembled into a matrix
${\bf \Phi} \equiv \underline{\phi} \cdot \underline{\tau}$ with
transformation law \be {\bf\Phi} (t, \vec {x}) \to M^{-1} (t, \vec
{x}) {\bf\Phi} (t, \vec {x}) M(t, \vec{x}) \ee and covariant
derivative \be D_\mu {\bf \Phi} \equiv \nabla_\mu {\bf {\Phi}} + g
[\bf{W}_\mu {\bf{\Phi}}- {\bf{\Phi}} \bf{W}_\mu ] \ee
\newpage
\noindent
or,
equivalently, \be D_\mu {\underline {\Phi}} \equiv \nabla_\mu
{\underline{\Phi}} + g \underline{W}_\mu \times {\underline{\Phi}} \
.\ee

In these expressions, $\mu = 0,1,2,3$. In what follows, we will ``work
in temporal gauge'', setting the time component of the gauge fields to
zero. Thus, the vector potential (or gauge potential) will be a three
vector and we will use the notation ${\vec \nabla} \times {\underline
{\vec W} } + g {\underline {\vec W}} \times {\underline {\vec W} } $
for the field strength; the expressions above should serve to clarify
whether the cross product is taken in internal space, in real space or
in both.

\section{Do-it-yourself magnetic monopoles}

This section is ``adapted'' ({\it i.e.} taken) from Coleman's
1974 Erice lectures \cite{C81}. He refers to this  as ``the monopole
hoax'', a joke to be played (or at least attempted) by a cunning
theorist on a gullible experimenter. In view of the number of cunning
experimenters in this audience I will refrain from comments and just
describe here how to build your own magnetic monopole.

1) Take a solenoid. It has to be very long and very thin so as to be
invisible; as Coleman says, {\it it helps if the solenoid is
many miles long and considerably thinner than a fermi (this is very
much a gedanken hoax).}

2) Put one end at the experimenter's laboratory,

3) hide the other end, and

4) turn on the current.

For a gullible theorist this may pass as a magnetic monopole, but of
course there is a way in which the solenoid could be detected: through
Aharonov-Bohm scattering.

The interference pattern in a double-slit experiment is shifted when a
solenoid is placed between the slits and the screen. Even if the
particle trajectories remain well outside the solenoid, their wave
functions $\psi_1, \ \psi_2$ acquire a phase $ \ exp ~[ i e \int {\vec
A}\cdot {\vec dl}] \ $ (with $e$ the electric charge of the particle
and the integral taken along the particle's path); if the paths are on
either side of the solenoid, the interference pattern changes because
the probability amplitude $|\psi_1 + \psi_2|^2$ becomes
\be | ~ e^{ie
\int_1 {\vec A}\cdot {\vec dl}} \psi_1 + e^{ie \int_2 {\vec A}\cdot
{\vec dl}} \psi_2|^2 = | \psi_1 + e^{ie \oint {\vec A}\cdot {\vec dl}
} \psi_2|^2 \ \ , \ee where $\oint {\vec A}\cdot {\vec dl} $, taken around
the solenoid, measures its magnetic flux.

Notice that, if the flux is an integer multiple of $2\pi /e$, the
solenoid becomes undetectable even quantum mechanically in our
gedanken experiment -- it is called a {\it Dirac string}.

The vector potential of a monopole whose Dirac string is along the
negative $z$-axis is given (in spherical coordinates $(r, \theta,
\varphi)$ centred on the monopole) by \be {\vec A_N}\cdot {\vec {dx}}
= {q_m \over 4\pi} (1 -\cos \theta) d\varphi, \qquad \qquad \theta
\neq \pi \ee giving rise to a radial magnetic flux\footnote { Note that
${\vec B}$ does not include the singular contribution from the Dirac
string.}  \be {\vec B} \cdot {\vec {dS}} = {q_m \over 4 \pi} \sin
\theta d\theta d\varphi \qquad\qquad {\rm or} \qquad\qquad {\vec B} =
{q_m \over 4\pi} {1 \over r^2} {\hat r} \ee where $q_m$ is the
magnetic charge.  The vector potential is singular on the Dirac
string, which is located at $\theta = \pi$, but regular everywhere
else.

Since the magnetic flux of the monopole is supplied by the Dirac
string, the condition that the flux through the string should be a
multiple of $2\pi/e$ gives rise to the famous Dirac quantization
condition for the magnetic charge $q_m$ of the monopole:
\be q_m ={ 2
\pi N \over e} \qquad {\rm or, \ reintroducing \ } \hbar \ {\rm and \ } c,
\qquad {e q_m \over 4\pi} = {N\over 2} (\hbar c) \ee Thus, the
existence of one monopole would be enough to force electric charge to
be quantized!

Note that we can use gauge invariance to change the position of the
Dirac string; an equivalent description of this monopole is given by
the vector potential \be {\vec A_S}\cdot {\vec dx} = - {q_m \over
4\pi} (1 +\cos \theta) d\varphi, \qquad \qquad \theta \neq 0 \ee
which is singular only at $\theta = 0$ (the gauge transformation
between $\vec A_N$ and $\vec A_S$ is singular in the position of both
the old and new strings, of course). You should not worry about these
singularities -- in the next section we will eliminate the Dirac
string altogether.

\section{The Wu-Yang construction of Dirac monopoles}

Electromagnetism is not a theory of gauge potentials {\it per se}, but
rather of equivalence classes of gauge potentials. This was exploited
by Wu and Yang \cite{WY75} to give a non-singular description of
magnetic monopoles by ``patching up'' vector potentials that are
regular in different regions, provided they are equivalent on the
overlaps.

Consider any sphere with non-zero radius surrounding the monopole. In
the northern hemisphere $ \ \theta \in [0, \pi/2 + \epsilon] \ $ take
${\vec A} = {\vec A}_N$ and in the southern hemisphere $ \ \theta \in
[\pi/2 - \epsilon, \pi]\ $ take ${\vec A} = {\vec A}_S$. In the
overlap region, $\pi/2 - \epsilon < \theta < \pi/2 + \epsilon$, the
two descriptions are related by a {\it regular} gauge transformation,
\be ({\vec A}_N - {\vec A}_S ) \cdot {\vec{dx}} = {\vec \nabla} \chi
\cdot {\vec {dx}} \qquad {\rm where} \ \chi (\varphi) = {q_m \over
2\pi}~ \varphi \ \ .  \ee This object has magnetic charge $q_m$, as
can be seen by computing the magnetic flux through the two hemispheres
(using Stokes' theorem). Note that, even though $\chi$ is not
singlevalued, $\vec \nabla \chi$ is, and therefore the gauge
transformation is well defined on the gauge potentials. Moreover,
singlevaluedness of the gauge transformation on the wave functions,
$e^{i e \chi}$, in the overlap equatorial region implies the Dirac
quantisation condition! $q_m \equiv \chi(2\pi) - \chi (0) = 2\pi N /e$.

We have eliminated the need for a Dirac string -- note that the only
singularity in this description is at the origin, $r=0$. But this must
be a real singularity because the energy of the monopole diverges as
$r \to 0$ due to the $1/r^2$ behaviour of the magnetic field,

\be E = \int d^3 x ~{1 \over 2} ~ ( |{\vec E}|^2 + |{\vec B}|^2 ) \sim
\int_0^\infty {dr \over r^2} \ee

Electromagnetism is not the only force in nature. There are the weak
and strong nuclear forces, and also gravity. We think that forces may
become unified into one kind of interaction (a Unified Theory) at high
energies\footnote{A special class are so--called Grand Unified
Theories, or GUTs, where the electromagnetic, weak and strong
interactions are described by a single simple group.}. When theorists
started to investigate possible unified theories they found a
surprise...

\section {'t Hooft--Polyakov monopoles}

One of the first attempts to unify the electromagnetic and weak
interactions was the $O(3)$ Georgi-Glashow model \cite{GG72} in which
the fundamental fields are a triplet of (real) scalars and a triplet of gauge
potentials.
\be
\eqalign{
&{\underline \phi} = (\phi^1,\phi^2,\phi^3) \cr
&{\underline {\vec W}} = ({\vec W}^1, {\vec W}^2, {\vec W}^3) \cr}
\ee

The most important aspect from our point of view is the energy, since
we are seeking to remove the divergence at $r=0$. I am making many
simplifying assumptions here (no time dependence, no electric fields),
and only writing the terms in the energy that are relevant for the
argument:

\be E =
\int d^3 x ~[ ({\vec \nabla} {\underline \phi} + g {\underline {\vec W}} \times
{\underline \phi} )^2 +
({\vec \nabla} \times {\underline {\vec W} } + g
{\underline
{\vec W}}
\times {\underline {\vec W} })^2 +
\lambda ({\underline \phi} \cdot {\underline \phi} - \eta^2)^2 ]~
\ee

First of all, the Georgi--Glashow model includes electromagnetism as a
subset: the configuration \be \eqalign{ &\phi^1 = \phi^2 = 0 \ ,
\qquad\qquad\quad \ \ \phi^3 = \eta \ , \cr &{\vec W^1} = {\vec W^2} =
0 \ , \qquad\qquad {\vec W^3 } = {\vec A} \ , \cr }\ee where $\vec A$
is any solution to Maxwell's equations, is also a solution of the full
non--abelian field equations of the Georgi--Glashow model. In
particular, the Wu-Yang (or Dirac) monopole is a solution.  However, it
is not a stable solution!

Indeed, the $1/r^2$ divergence in the monopole energy is now coming
from $\vec{\nabla} \times \vec{W^3}~$ in the $~\int ({\vec \nabla}
\times {\vec W^3} + g {{\vec W^1}} \times {{\vec W^2 }})^2~$ term and
could be controlled if $\vec W^1$ and $\vec W^2$ acquired non-zero
values $\sim 1/r$. This is consistent with the fact that $\vec{W}^\pm
\equiv (\vec W^1 \mp i \vec W^2) /\sqrt 2 $ are charged fields (the
{\it W--bosons}) with charge $\pm g$ respectively and a magnetic
moment $ i g {\vec W}^- \times {\vec W}^+$ which couples to the $\vec
W^3$ magnetic field, so their presence can reduce the magnetic energy. On
the other hand, such ``W-condensation'' has two immediate effects: ~one
is an increase in energy coming from the new, non--zero $ ({\vec
\nabla} \times {\vec W^{1,2}}\pm g {{\vec W^3}} \times {{\vec W^{2,1}
}})^2$ terms; the other is that the scalar gradients
$D_\mu \phi^{1,2} \sim
g {\underline {\vec W}^{2,1}} \times {\underline {\phi}^3}$ now
diverge as $1/r$. However this problem is
eliminated if $\phi^3 \sim r$ as $r \to
0$. The condition $\underline{\phi}(r=0) = 0$ imposes a penalty in
energy from the $\int \lambda ({\underline {\phi}} \cdot {\underline
{\phi}} - \eta^2)^2$ term, but this is {\it finite} -- thus, the
result is always energetically favourable to the singular abelian monopole
that we started with.

In the case when magnetic charge is two Dirac units, $q_m = 4\pi/g$
(see below for an
explanation of this condition),
this instability leads
to the 't Hooft--Polyakov monopole \cite{HP74}, a spherically symmetric
configuration describing a non-singular magnetic monopole of finite
mass \cite{CM},

\be
\eqalign{
&\phi^1 = \phi^2 = 0 \ , \qquad\qquad\qquad\qquad\qquad\qquad\qquad
 \phi^3 = \eta \rho(r)\cr
&{\vec W^1 - i\vec W^2} = {1 \over g} {f(r) \over r} e^{i \varphi} (
{\hat{\varphi} \over \sin\theta}  - i{\hat \theta}) ,
\qquad\qquad {\vec W^3 } = {1\over g} (1 -\cos
\theta) {1 \over r \sin\theta} {\hat \varphi}
\ ,
\cr }
\ee
with $f(r) \to 1$, $f'(r) \sim - r$ and $\rho(r) \sim r$ as $r \to 0$ and
$f(r) \to 0 , \  \rho(r) \to 1$ as $r \to \infty$.

Note that only the small $r$ behaviour of the fields has changed; in
particular, the magnetic charge of the monopole remains the
same. After a (singular) gauge transformation it reduces to the more
familiar form \cite{HP74}
\be \eqalign{ {\underline
{\phi} } &= \rho(r) {\hat r} \cr {\underline {\vec W}} &= {1 \over g}
(f(r) -1) {\hat r} \times {\vec \nabla} {\hat r} \ \ ,\cr} \ee which
shows that the 't Hooft--Polyakov monopole is a topological defect
(usually called a {\it hedgehog} because of the way the scalar field
points radially outwards). The zero value of the scalar field at the
origin $r=0$ is forced by the non-trivial winding of the scalar field.

It remains to explain why the restriction to two units of magnetic
charge.  In the Wu-Yang construction we started with a sphere of
non--zero radius, say $R$, divided into two hemispheres overlapping at
the equator. If the monopole has $N$ units of magnetic charge the
gauge transformation in the overlap region is a phase rotation by
$2\pi N$.  W--condensation replaces the singularity at the origin by
an everywhere regular core. Since nothing changes outside the core,
the patching condition for the 't Hooft-- Polyakov monopole remains a
$2\pi N$ rotation for all $R>r_{\rm core}$. But this cannot be true inside
the core: since there is no singularity,
the gauge transformation must also change continuously so that it
becomes the identity when we reach $r=0$. If the gauge group is
$U(1)$, this is simply not possible, and all monopoles are
singular. But in $SU(2)$, a $2\pi$ rotation is not continuously
connected to the identity whereas a $4\pi$ rotation is!  Thus, only
monopoles with even $N$ can be non-singular. Of all these, it turns
out that only $N=2$ remains spherically symmetric after
W--condensation.

The existence of magnetic monopoles is a very generic prediction for a
large class of theories containing electromagnetism. Moreover, they
should be produced in large numbers in the early Universe
\cite{P79}. The fact that we do not observe those monopoles is a
serious challenge to cosmologists, and has become known as {\it the
monopole problem}.

\section{Magnetic monopoles in the Weinberg--Salam model; electroweak
strings and dumbells}

The standard  model of electroweak interactions has a $SU(2)
\times U(1)$ gauge symmetry, corresponding to weak isospin and hypercharge
respectively.
Its bosonic sector comprises a neutral
scalar field
$\phi^0$,
a charged scalar field
$\phi^+$, and the vector potentials corresponding to the  massless photon
$\vec A$ and three massive vector bosons: the charged W--bosons
(${\vec W}^\pm$) and the  neutral $\vec Z$.
The fermionic sector consists of the  three  families of quarks and
leptons
\be
     \pmatrix{\nu_e \cr e \cr u\cr d\cr} \qquad
     \pmatrix{\nu_\mu \cr \mu \cr c\cr s\cr}\qquad
            \pmatrix{\nu_\tau \cr \tau \cr t\cr b\cr}
\ee

In the Weinberg-Salam model, the electromagnetic and $Z$-fields are
combinations of the $SU(2)$ and $U(1)$ gauge potentials
($\underline{\vec{W}}$ and $\vec Y$ respectively):
\begin{equation}
{\vec Z} \equiv \cos\theta_w{\vec W}^3 - \sin\theta_w {\vec Y} \ ,
\ \ \ \
{\vec A} \equiv \sin\theta_w{\vec W}^3+ \cos\theta_w {\vec Y} \ ,
\end{equation}
where $\theta_w $ is called the weak mixing, or Weinberg, angle. Its
measured value is $\sin^2\theta_w \approx 0.23$.

In this case, the field that satisfies Maxwells equations at low
energy is $\vec A$ and, being massless, it is the only vector
potential that can give rise to long-range electric and magnetic
fields and thus magnetic monopoles. Its configuration far from the
monopole will be exactly like what was discussed in sections 3 and
4. Very close to the monopole, though, we would expect other fields to
condense due to the intense magnetic field, changing the core
structure.

But there is a problem. Note that, since the electromagnetic field has
a hypercharge component and hypercharge is an abelian field,
isolated magnetic monopoles are always singular
at the origin. We are back to square one!

In order to have monopoles with regular cores one has  to embed the
$SU(2) \times U(1) $ symmetry of the Weinberg--Salam model into larger
symmetry groups. We already mentioned Grand Unified Theories (GUTs,
for short), where one simple group not only contains the electroweak
interaction, but also the strong interaction. These monopoles are very
heavy, because the unification of these forces occurs at very high
energies ($\sim 10^{16}$ GeV) and the fields that condense at the core
are very massive. Far too heavy to be produced in a particle
accelerator.

But there are also lighter magnetic monopoles in the Weinberg--Salam
model: they occur as monopole--antimonopole pairs connected by a
vortex (the vortex carries magnetic flux of the $Z$-boson, and it is
usually called a $Z$--{\it string} or an electroweak string). Such
configurations were called {\it dumbells} by Nambu, who first
considered them in 1977 \cite{N77}. The dumbell is rotating to avoid
longitudinal collapse, and its mass is estimated at a few TeV.

Their internal structure is rather interesting. The $SU(2)$ fields are
those of a 't Hooft--Polyakov monopole--antimonopole pair, while the
hypercharge $U(1)$ field configuration resembles that of a solenoid
joining the monopole and antimonopole. As a result, the combination
inside the solenoid is precisely the magnetic part of the $Z$ field,
whereas the magnetic field that emanates from the solenoid ends is the
massless electromagnetic field, {\it and there are no singularities
anywhere}.

In some respects, the structure of the $Z$--string is similar to that
of a magnetic vortex in an Abrikosov lattice that appears in a type II
superconductor subjected to an external magnetic field. In the
Weinberg--Salam model, the role of the vector potential is taken by
the $Z$--field, and the order parameter is the neutral Higgs field
$\phi^0$. But there is a very important difference: electroweak
strings are non--topological, and therefore not necessarily stable.

A review of electroweak strings where the stability issue is discussed
in some detail can be found in \cite{AV99}. Stability depends on the
value of the Weinberg angle and the masses of the various
fields.
In the physical range of these parameters, it is found
that infinitely long, straight, bare strings are classically unstable
(they are only stable for $\theta_w \sim \pi/2$ and $m_{\rm
Higgs} < m_Z$).
On the other hand, stability improves for short segments and in the
presence of magnetic fields.  Finally, the strings are superconducting
and fermion modes on the string might also stabilize them, although
this is still under discussion.  The stability of dumbells
remains an open question.

Even in the region of parameter space where strings are (classically)
stable, the fact that stability is not topological immediately raises
the question of whether a network of strings would form in a phase
transition via the Kibble mechanism. We now turn to this question, and
we will focus on the (unphysical) limit of the Weinberg--Salam model
in which the $SU(2)$ symmetry becomes global, while keeping the
hypercharge $U(1)$ symmetry local. In this limit, known as the
semilocal model, electroweak strings are classically stable.

\section{Semilocal strings}

The semilocal model is obtained when the complex scalar field $\phi$
in the Abelian Higgs (or Landau--Ginzburg) model is replaced by an
$SU(2)$ doublet of complex fields $(\phi,\psi)$. \footnote{Similar systems
have been considered in the condensed matter literature
\cite{BK87,V84}.}

It is also (the bosonic sector of) the Weinberg--Salam model in the
limit in which the $SU(2)$ gauge coupling is set to zero; in this
limit, the $W$-bosons and the photon decouple and the symmetry is
$SU(2)_{\rm global} \times U(1)_{\rm local}$. The only gauge field is
the neutral $Z$ boson, which coincides with the hypercharge field. The
model has vortices ({\it semilocal strings}) whose properties are
intermediate between electroweak strings and
Abrikosov--Nielsen--Olesen vortices (see \cite{AV99}).

The energy per
unit length of cylindrically symmetric configurations  is \be E = \int d^2 x [
|({\vec \nabla} - iq {\vec Y} ) \phi |^2 + |({\vec \nabla} - iq {\vec
Y} ) \psi |^2 + {1 \over 2} ({\vec \nabla} \times {\vec Y} )^2 +
\lambda (\phi^2 + \psi^2 -{\eta^2\over 2})^2 ] \ee and it turns out that, even
though the vacuum manifold is a three-sphere, \be \phi^2 + \psi^2 =
{\eta^2 \over 2}\ee which is simply connected, $\pi_1(S^3) = 1$, there {\it
are} stable strings if the scalar mass is smaller than the vector
mass. These strings are not only stable classically, they are also
stable to semiclassical tunnelling and to breaking by monopole
pairs. If $m_{\rm scalar} > m_{\rm vector}$ the strings are
classically unstable, and if the masses are equal there is a
two-parameter family of configurations with the same energy where the
quantised magnetic flux $2\pi/q$ spreads over an arbitrarily large
core width.

Semilocal strings, like electroweak strings, can have open ends; but
the monopoles at the ends are global monopoles
and have a
long--range interaction.

The stability of semilocal strings is well understood in terms of the
competition between gradient energy and potential energy. The gauge
field can compensate gradients in the (complex) phase of either of the
two scalar fields, but it can only compensate both gradients {\it
simultaneously} if there is a correlation between the phases of the
two scalars; this correlation is present in the string solution, which
optimizes gradient energy, but at the expense of a large concentration
of potential energy at the core. If $\lambda/q^2$ is very large, it
becomes energetically favourable to break the phase correlation in
order to reduce potential energy, and the string is destroyed. If
$\lambda$ is small, the cost in gradient energy to dissolve the string
becomes too high and the string is stable.

Since the strings are non-topological, the question immediately arises
as to whether they would form at all in a phase transition. In
particular, stability depends on the mass ratio $m_{\rm scalar} /
m_{\rm vector}$ and we would expect the formation rates to reflect
this dependence.

We now turn to a first attempt to answer these questions through
numerical simulations.  Note that setting $\psi = 0$ in the semilocal
model gives the Abelian Higgs (or Landau--Ginzburg) model, thus making it
possible to compare the rates of formation of semilocal strings to
those of topological strings in a similar environment.  The
conclusion seems to be that semilocal strings do form, and
in some cases with number densities comparable to those of their
topological counterparts.

\begin{figure*}
\centering
\leavevmode\epsfysize=7cm \epsfbox{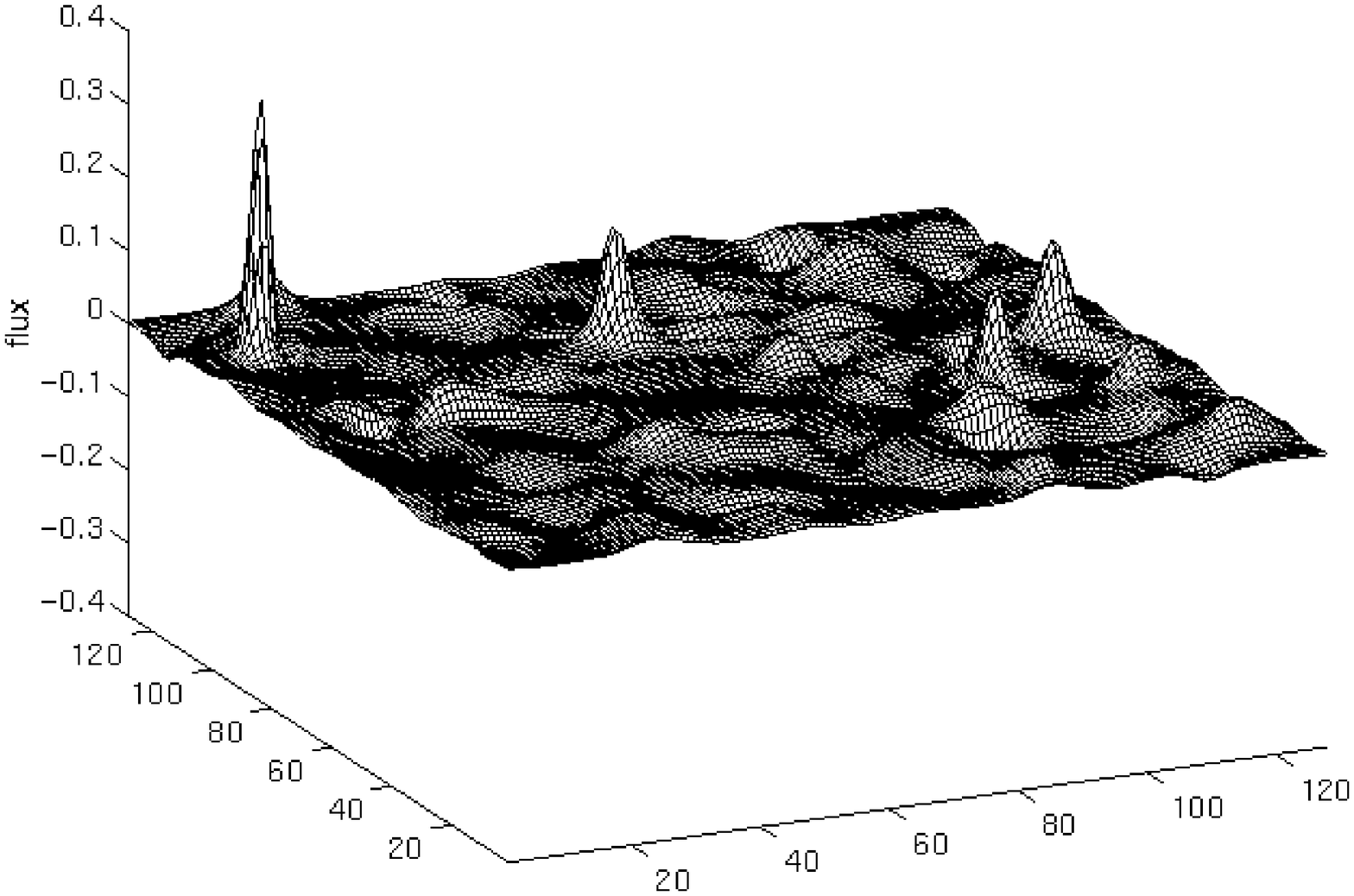}\\
\leavevmode\epsfysize=7cm \epsfbox{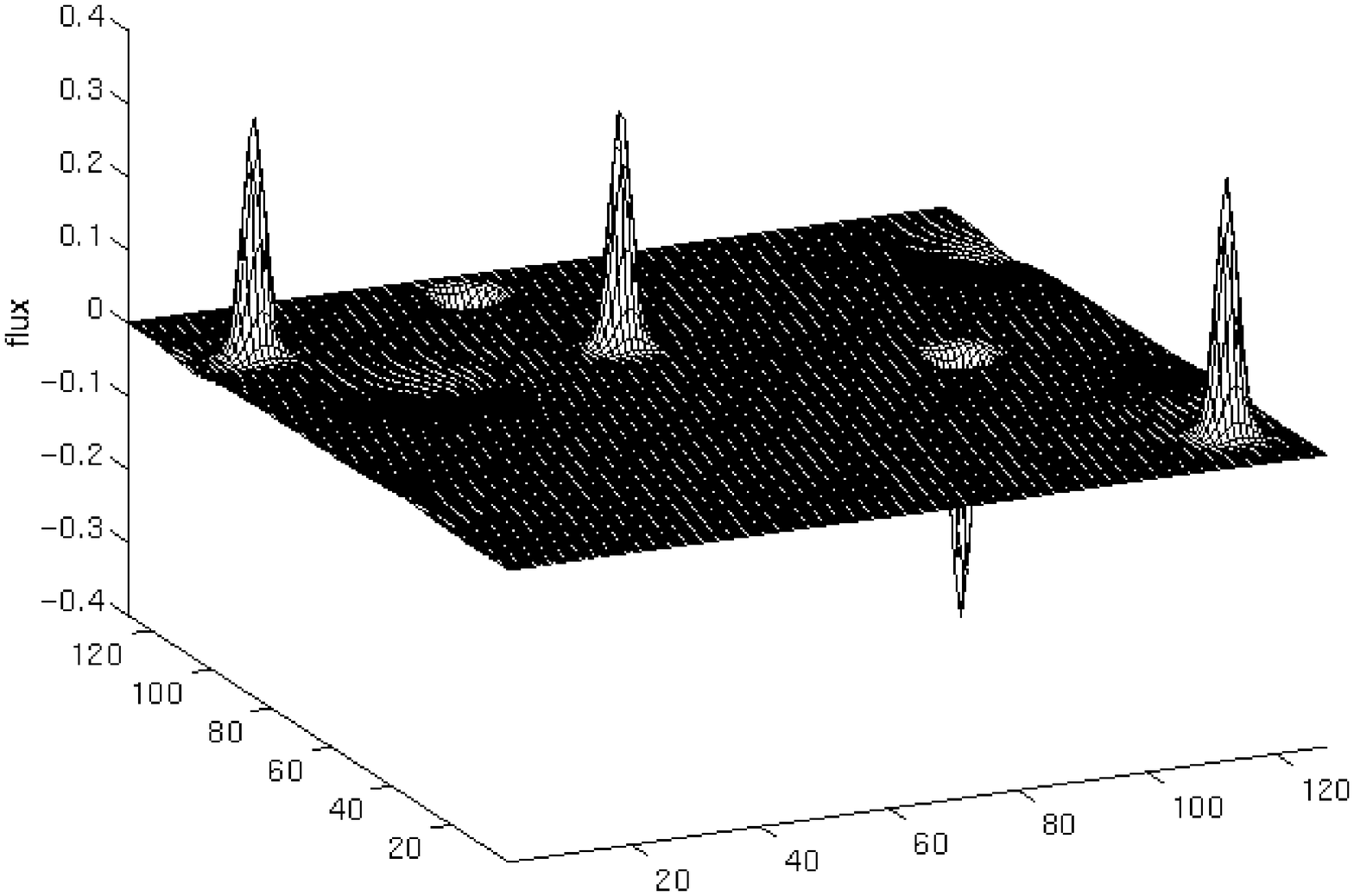}\\
\caption[semi1]{\label{semi1} The flux tube structure in a
 two-dimensional semilocal string
 simulation with $\beta = 0.05$. The upper panel ($t=0$) shows
 the initial condition
 after the process described in the text. The lower panel shows the
 configuration resolved into five flux tubes by a short period of
 dynamical evolution ($t=100$). These flux tubes are semilocal vortices.
{Note the different numbers
of upward and downward
 pointing flux tubes,
 despite the zero net flux boundary condition. The missing flux
 resides in the smaller `nodules', made long-lived by the numerical
viscosity; the expansion of the universe could have a similar effect and
preserve these
 `skyrmionic' configurations \cite{BB}.}}
\end{figure*}

\section{Numerical simulations of semilocal string formation}

\begin{figure}[t]
\centering
\leavevmode\epsfysize=5cm \epsfbox{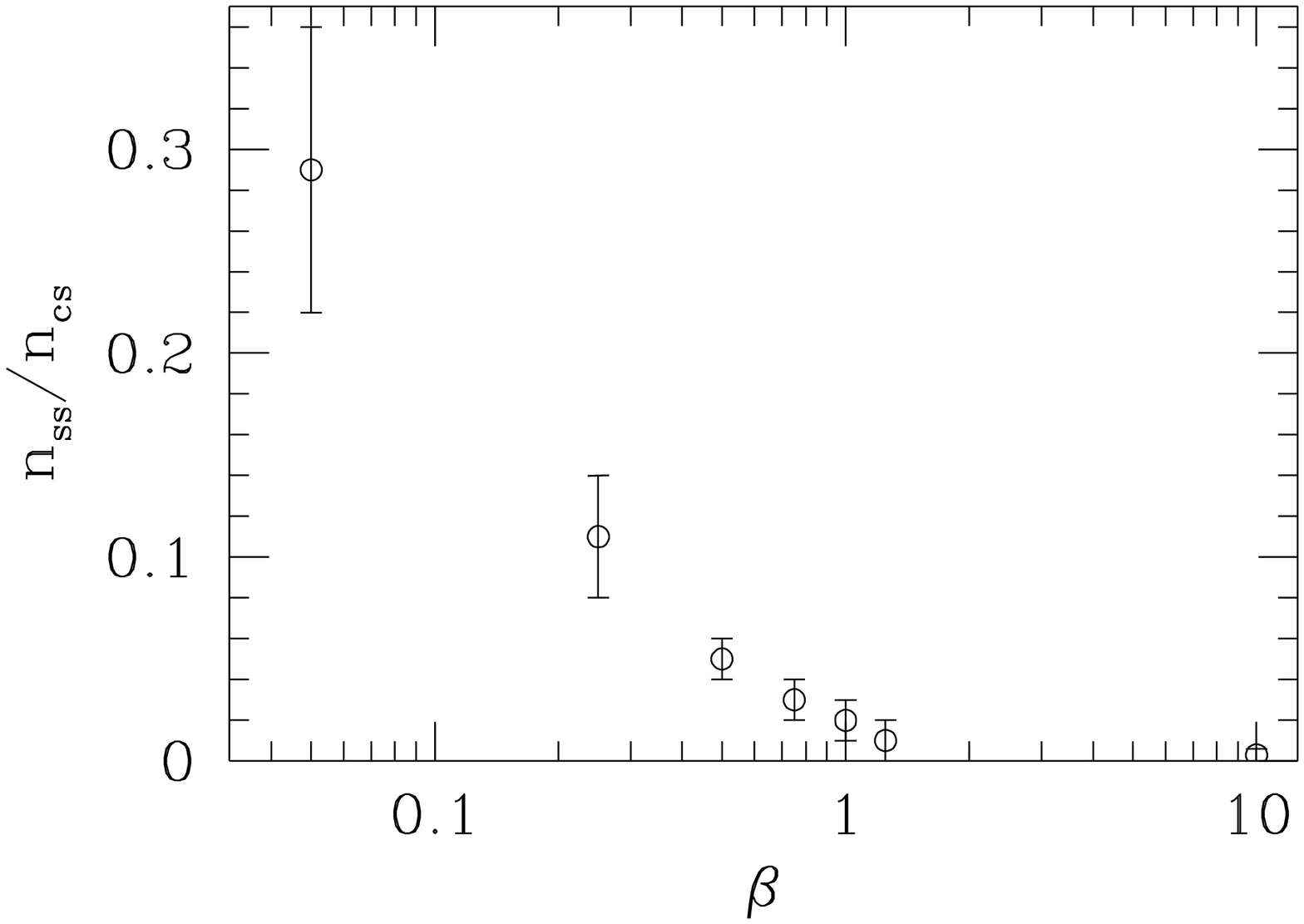}\\
\caption[semi6]{\label{semi6} The ratio of lengths of semilocal and cosmic
strings.}
\end{figure}

For details of the simulations we refer the reader to
\cite{ABL97,ABL98}. Here we will just point out the
main features and results, summarized in figure \ref{semi6}.

Space is discretized into a lattice with periodic boundary conditions.
The equations of motion are solved numerically using a standard
staggered leapfrog method, and a dissipation term ($\eta \dot{\phi}, \
\eta \dot\psi$ or $\eta \dot{Y}_i$) is added to each equation to
reduce the relaxation time. A range of strengths of dissipation was
tested, and it did not significantly affect the number densities
obtained; the simulations displayed in the figures all have have $\eta
= 0.5$.  Note that in an expanding Universe the expansion rate would
act as a sort of viscosity, though $\eta$ would typically not be
constant.  However, this is not meant to be a cosmological simulation
since spacetime is flat.

We work in temporal gauge. Then Gauss' law becomes a constraint
 derived from the gauge choice $Y_0=0$, and is used to test the
 stability of the code.

The inverse vector mass is taken as the
unit of length and time, and $\eta^2$ as the unit of energy. In these
units the dynamics is governed by a single parameter, \be \beta =
m_{\rm scalar}^2/m_{\rm vector}^2 = {2\lambda \over q^2} \ee which also
determines the
stability of the straight string solutions: they are stable if $\beta
<1$ and unstable if $\beta >1$. \footnote{ $\beta$ is also the parameter
that distinguishes type I ($\beta <1$) from type II ($\beta >1$)
superconductors.}

\begin{figure}
\centering
\leavevmode\epsfysize=6.5cm{\epsfbox{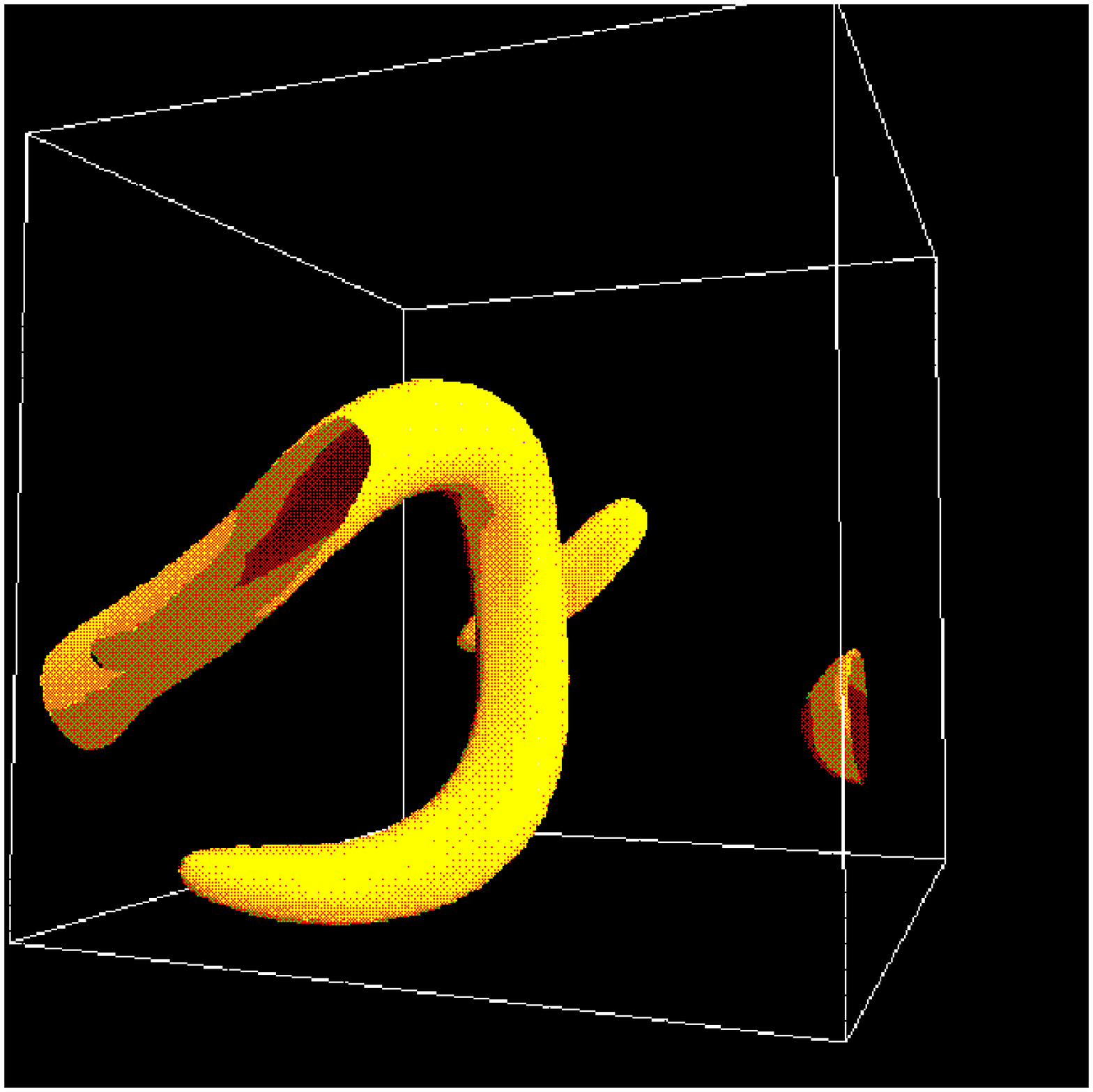}}\\
\leavevmode\epsfysize=6.5cm{\epsfbox{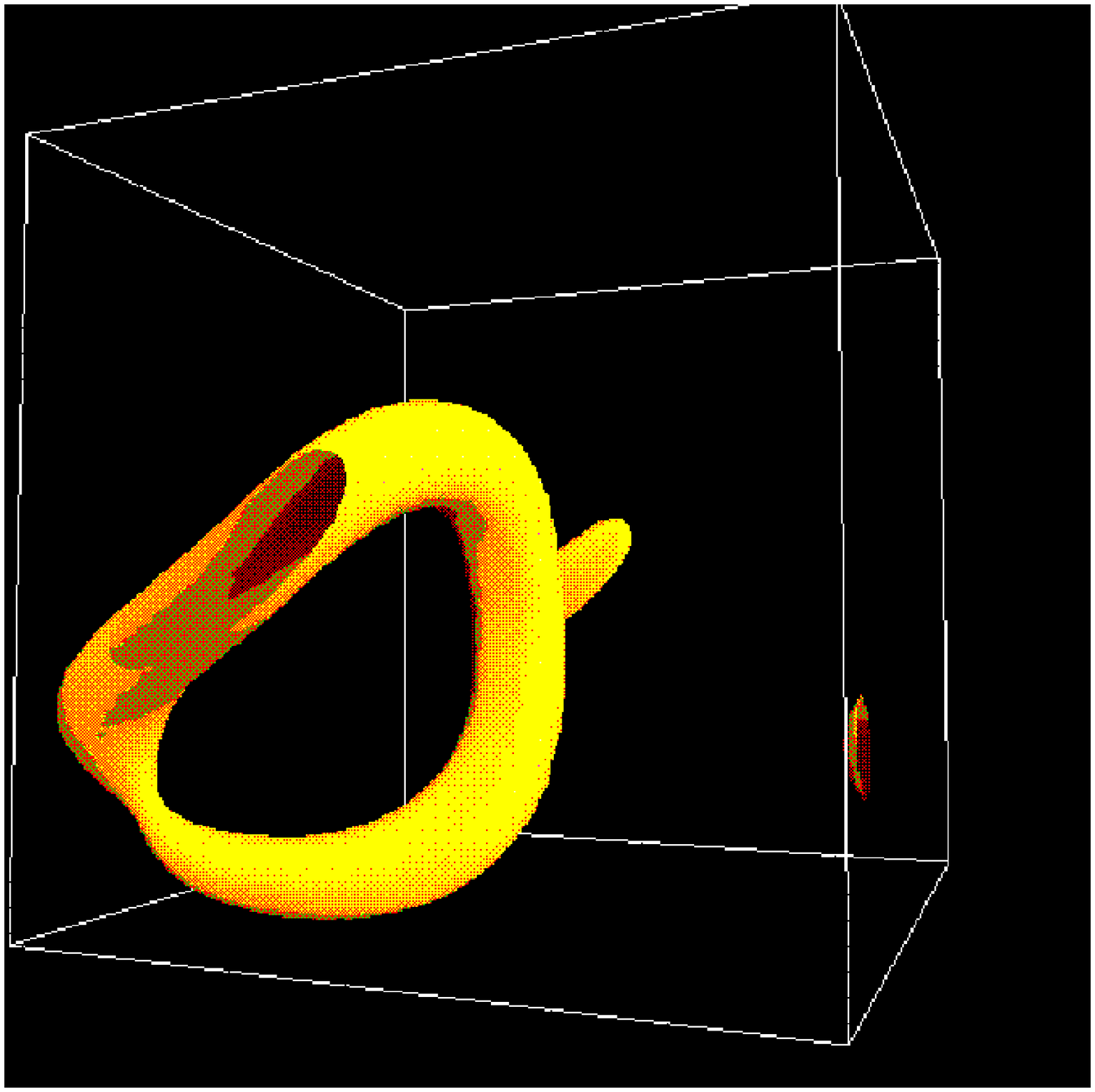}}\\
\caption[loopformation]{\label{loopformation} Loop formation from
semilocal string segments.  The figure shows two snapshots, at $t =
70$ and $t = 80$, of a $64^3$ numerical simulation of a network of
semilocal strings with $\beta = 0.05$, where the ends of an open
segment of string join up to form a closed loop. Subsequently the
loops seem to behave like those of topological cosmic string,
contracting and disappearing.}
\end{figure}

The number density of defects is estimated by a multi-step process.
The initial conditions are obtained with a generalization to
non-topological strings of what is known to cosmologists as the
Vachaspati--Vilenkin algorithm \cite{VV84}, followed by a short period
of dynamical evolution.

$\bullet$ First, we generate a random initial configuration for the
scalar fields drawn from the vacuum manifold, which is not
discretised.  If space is a grid of dimension $N^3$, the correlation
length is chosen to be some number $p$ of grid points ($p=16$ in \cite
{ABL97,ABL98}; the size of the lattice is either $N=64$ or $N=256$).
To obtain a reasonably smooth configuration for the scalar fields, we
assign random vacuum values on a $(N/p)^3$ subgrid and interpolate the
scalar field smoothly onto the full grid.

$\bullet$ We then find the gauge field configuration that minimizes
the energy in this fixed scalar background.

Two dimensional test simulations have shown that the energy minimization
is redundant, since the early stages of dynamical evolution carry out
this role anyway; for simplicity, we used in practice a gauge field
configuration which was close, but not equal, to the real minimum.

$\bullet$ An example of the initial conditions generated with this
algorithm can be seen in Figure \ref{semi1} in the case of a
two-dimensional toy model with translational invariance in one
dimension, say $z$.
The plot shows
magnetic field on the $x-y$ plane, which has
a complicated flux structure with extrema
of different values (top panel of Fig. \ref{semi1}), and it is far
from clear which of these, if any, might evolve to form semilocal
vortices; in order to resolve this ambiguity, the
initial configurations are
evolved forward in time with zero initial velocities for the fields. After
a short transient, in the unstable regime
$\beta > 1$ the flux quickly dissipates leaving no strings. By
contrast, in the stable regime $\beta < 1$ stringlike features emerge
when configurations in the ``basin of attraction''of the semilocal
string relax unambiguously into vortices (bottom panel of
Fig. \ref{semi1}). We only count vortices after this relaxation process.

\begin{figure}
\centering
\leavevmode\epsfysize=6.5cm{\epsfbox{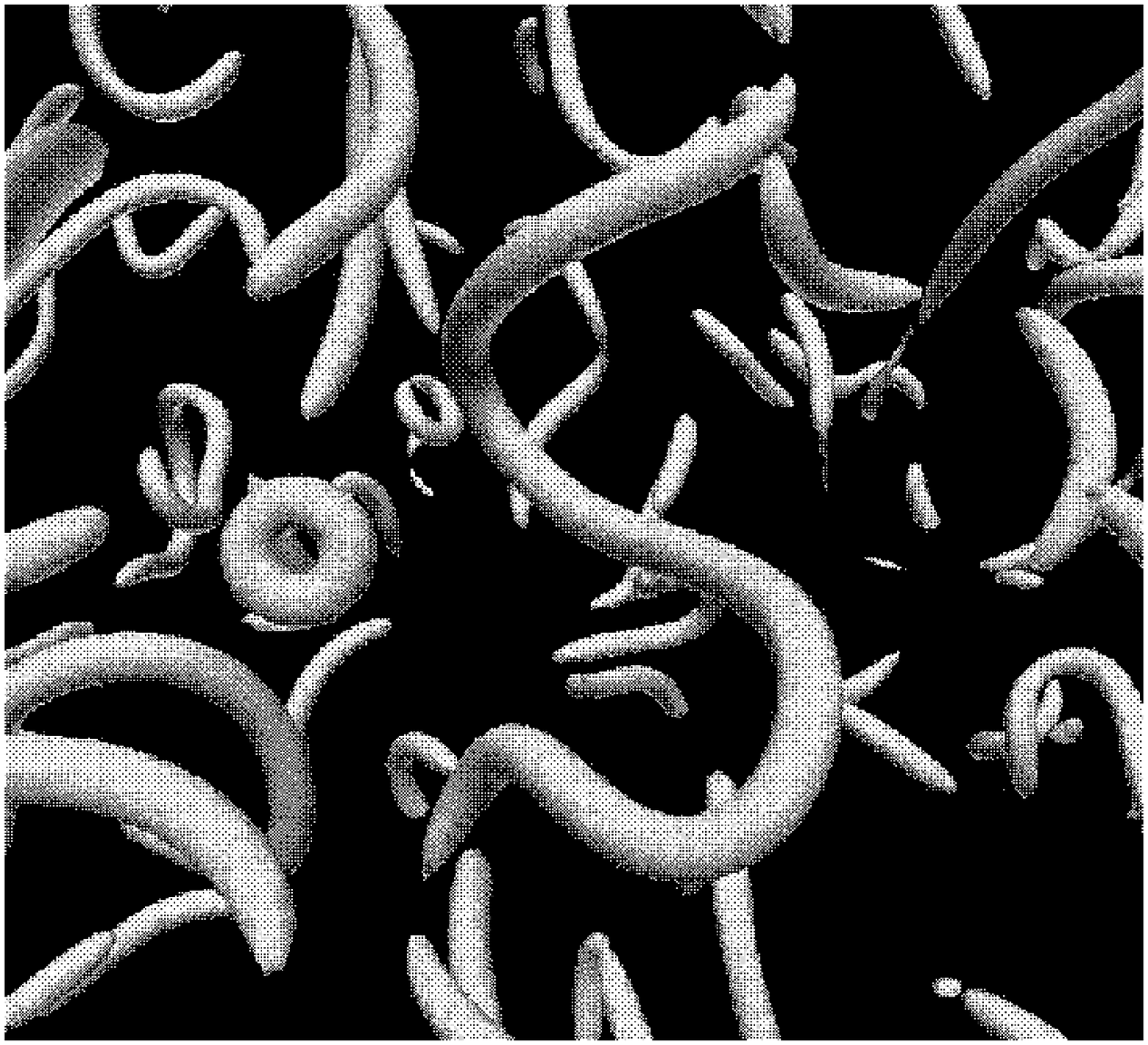}}\\ 
\vspace*{5pt}
\leavevmode\epsfysize=6.5cm{\epsfbox{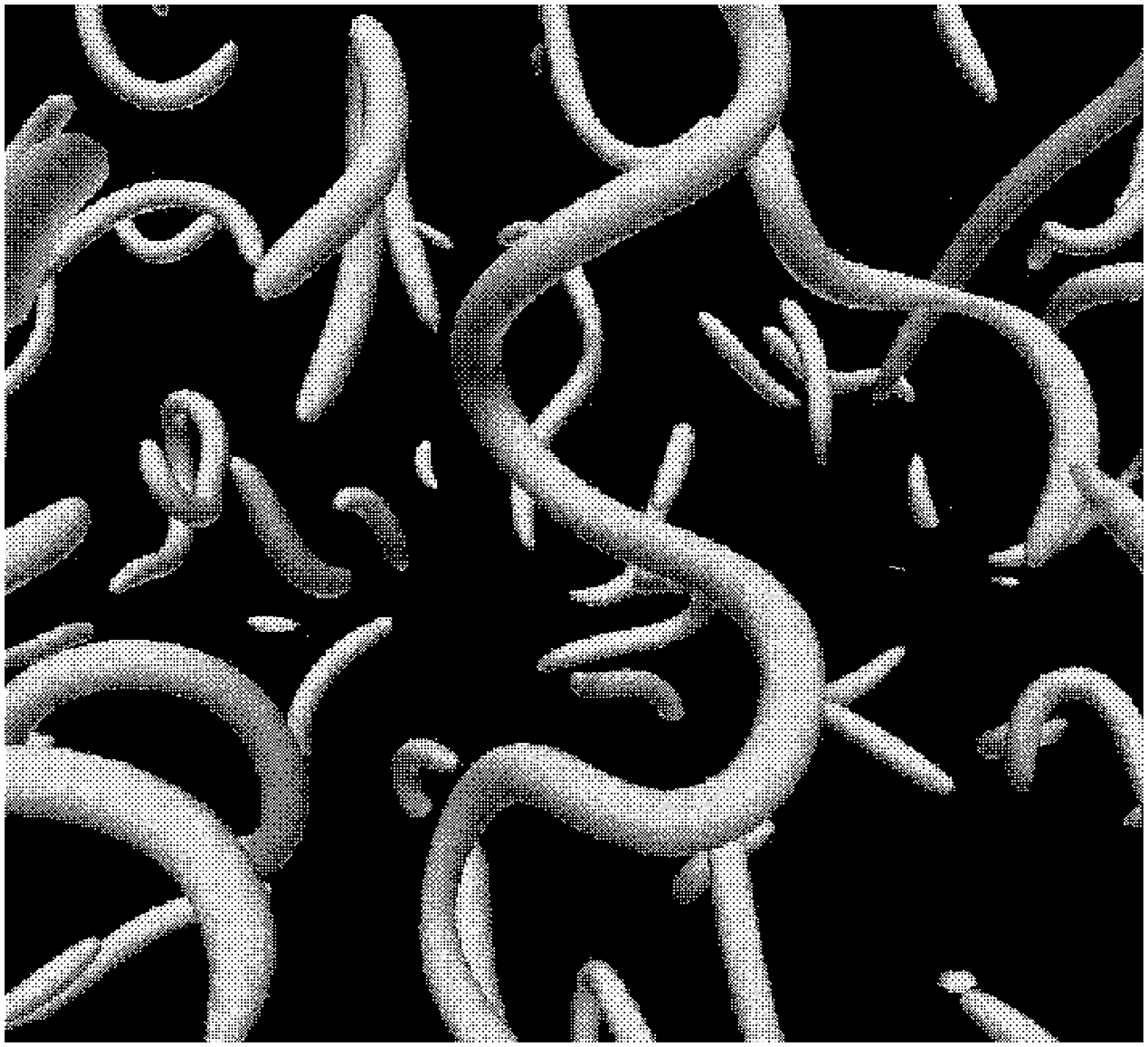}}\\ 
\vspace*{5pt}
\caption[semi4]{\label{semi4} {\it The growth of string segments to
form longer strings}. The figure shows two snapshots, at time $t=60$
and $t=70$ of a large $256^3$ numerical simulation of a network of
semilocal strings with $\beta = 0.05$. Note several joinings of string
segments, e.g.~two separate joinings on the long central string, and
the disappearance of some loops. The different apparent thickness of
strings is entirely an effect of perspective. The simulation was
performed on the Cray T3E at the National Energy Research Scientific
Computing Center (NERSC) in Berkeley.}
\end{figure}

$\bullet$ One important point is that strings are always identified
with the location of magnetic flux tubes, rather than by the zeroes of
the scalar field.  Figure \ref{semi4} shows two snapshots of a simulation
on a $256^3$ lattice.
In order to make the figure, and also to compute the number density,
we need to set a magnetic flux threshold -- the strings that can be
seen in that figure are made of those points in which the magnetic field
exceeds half the maximum value in an Abrikosov-Nielsen-Olesen vortex.
After the short transient, this threshold can be modified without
significantly affecting the results, which are ratios of semilocal to
topological string number densities (shown in figure \ref{semi6}).
The error bars include, among other things, this threshold dependence
as well as the dispersion between runs.

\begin{figure}
\centering
\leavevmode\epsfysize=5cm \epsfbox{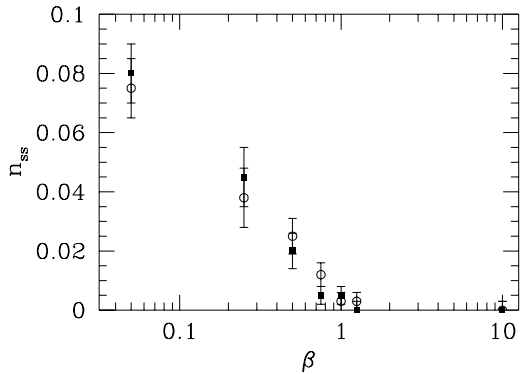}\\
\caption[semi2]{\label{semi2} The number of semilocal strings formed
per initial two-dimensional correlation volume in a toy model with
translational invariance in one direction. Each point is an average
over ten simulations.  Squares indicate that the vacuum initial
conditions described in the text were used, while open circles
indicate that non-vacuum (more thermal) initial conditions were used.}
\end{figure}

Since the initial conditions are somewhat artificial, the results were
checked against various other choices of initial conditions, in
particular different initial conditions for the gauge field and also
initial conditions for the scalar field closer to a thermal
environment (see Fig. \ref{semi2}).  However, it cannot be
sufficiently stressed that it is not a realistic thermal simulation of the
phase transition. There is no thermal noise and all the initial
conditions in \cite{ABL97,ABL98} had zero initial velocities for the
fields -- initial conditions with non-zero field momenta have not yet
been investigated --. What the simulations show is that the rate of
formation of these non-topological vortices is not only non--zero but
in certain cases it could  even be comparable to that of their
topological counterparts. Obviously the details of the transition can
and should now be investigated in full.

\section{Discussion and outlook}

The physical mechanism behind the observed string formation by
accretion of magnetic field in Figure \ref{semi1} or by growth of
string segments in Figures \ref{loopformation} and \ref{semi4} is not
very different from the formation of Abrikosov lattices in a type II
superconductor, even though in a system such as the early Universe it
makes no sense to talk about external magnetic fields. Because of the
$Y^2\Phi^2$ term in the energy, the magnetic field does not like to
coexist with the superconducting ($\phi \neq 0$) phase of the scalars.

In the topological strings case one usually argues that the non-zero
winding in the phase of the scalar field forces a zero of the Higgs,
and magnetic flux gathers there in a vortex. Conversely, what we are
observing in the semilocal model is that if there is a sufficiently
large concentration of magnetic flux in a small region, for instance
near the end of a string segment or maybe due to a fluctuation, a line
of zeroes of the Higgs can develop there. Small segments of
string, on the other hand, will tend to contract and disappear.

It seems clear that these results should extend to electroweak strings
with $g \neq 0$ as long as we remain in the region of parameters where
the strings are classically stable \footnote{Note that the monopoles
at the ends of semilocal strings are global, whereas those at $g \neq
0$ will have finite size cores. However, the core size grows as $1/g$
and, for sufficiently small $g$, it may be larger than the average
distance between string segments, causing the strings to grow}.  In
the region of stability, we expect a non-zero density of electroweak
vortices to form in a phase transition. Preliminary results seem to
confirm this picture \cite{S99}, so obviously this problem deserves
attention.

We may be a long way from understanding the formation of magnetic
monopoles in a phase transition, but it is possible that particle
accelerators will show signatures of monopole-antimonopole pairs in
the not too distant future.  If Nambu's prediction is correct,
dumbells could be the first soliton-like objects in the standard model
of particle physics to be observed.

\section{Acknowledgments}
I am very grateful to Henri Godfrin and Yuriy Bunkov for organizing
such a stimulating school, and to Brigitte Rousset and the Les Houches
secretariat for their help at various points. Thanks also to my
collaborators Julian Borrill and Andrew Liddle, and to Paul Saffin for
a preview of his results on electroweak string formation.

\end{document}